\documentclass[aps,pra,showkeys,superscriptaddress]{revtex4-2}
\usepackage{array}
\usepackage{booktabs}
\usepackage{dcolumn}
\usepackage{amsmath}
\usepackage{amsfonts}
\usepackage{amssymb}
\usepackage{graphicx,color}
\usepackage{float}
\usepackage[colorlinks={true}]{hyperref}
\hypersetup{colorlinks=true,linkcolor=red,citecolor=cyan,urlcolor=blue}
\usepackage{graphicx}
\usepackage{subfigure}
\usepackage{graphicx}% Include figure files
\usepackage{dcolumn}% Align table columns on decimal point
\usepackage{bm}% bold math
\usepackage{orcidlink}
\usepackage{pstricks}
\usepackage{float}

\usepackage{pgfplots}
\pgfplotsset{compat=1.16}
\usepackage{siunitx}
\usepackage{tikz}
\usepackage{tikz-cd}
\usepackage{pgfplots}
\usepackage{overpic}

\def\be{\begin{equation}}
	\def\ee{\end{equation}}
\def\bea{\begin{eqnarray}}
	\def\eea{\end{eqnarray}}

\usepackage[absolute,overlay]{textpos} % Package to position elements absolutely

\begin{document}
	\title{The Limits of Quantum Information Scrambling}
	\author{M. Y. Abd-Rabbou\orcidlink{0000-0003-3197-4724}}\email{m.elmalky@azhar.edu.eg}
	\affiliation{School of Physics, University of Chinese Academy of Science, Yuquan Road 19A, Beijing, 100049, China}
	\affiliation{Mathematics Department, Faculty of Science, Al-Azhar University, Nassr City 11884, Cairo, Egypt}
	\author{Ahmed A. Zahia \orcidlink{0009-0009-6693-5080}
	} \email{ahmed.zahia@fsc.bu.edu.eg}
	\affiliation{Department of Mathematics, Faculty of Science, Benha University, Benha, Egypt}
	\author{ Atta ur Rahman \orcidlink{0000-0001-7058-5671}
	}
	\email{attazaib5711@gmail.com}
	\affiliation{School of Physics, University of Chinese Academy of Science, Yuquan Road 19A, Beijing, 100049, China}
	
	\author{ Cong‑Feng Qiao \orcidlink{0000-0002-9174-7307}
	}
	\email{qiaocf@ucas.ac.cn}
	\affiliation{School of Physics, University of Chinese Academy of Science, Yuquan Road 19A, Beijing, 100049, China}
	\affiliation{Key Laboratory of Vacuum Physics, University of Chinese Academy of Sciences, Beijing 100049, China}

\begin{abstract}
Quantum Information scrambling (QI-scrambling) is a pivotal area of inquiry within the study of quantum many-body systems. This research derives mathematical upper and lower bounds for the scrambling rate by applying the Maligranda inequality. Our results indicate that the upper bounds, lower bounds, and scrambling rates coincide precisely when local operators exhibit to be unitary-Hermitian. Crucially, the convergence or divergence of these upper and lower bounds relative to the scrambling rate is contingent upon the system's initial state. The spin-star model to validate this theoretical framework is investigated, considering thermal and pure initial states. The implantation of the ancilla or external qubit aligns the scrambling rate with the established bounds. The upper and lower bounds may diverge from the scrambling rate based on the system's initial state when both local operators are multi-quit systems. The scrambling rate found grows with the increase of the qubit number in local operators.
\end{abstract}
\maketitle
 
\section{Introduction}
The study of quantum information dynamics in many-body systems poses a profound enigma: how does quantum information encoded in the initial state spread across the entire system during temporal evolution? This inquiry is pivotal for advancing comprehension of such systems’ fundamental behavior \cite{lewis2019dynamics}. Intriguing parallels emerge with the black hole information paradox, initially conceptualized in relativistic contexts \cite{PhysRevD.73.104027}. Contemporary research posits that in non-classical many-body systems, information within initial classical correlations becomes non-locally entangled and diffusely distributed during temporal evolution \cite{bhattacharyya2022quantum, shenker2014black, sekino2008fast}. While recovering such information via classical means may eventually become feasible, this would demand comprehensive global measurements—a process tied to quantum information scrambling \cite{PhysRevA.94.040302}. Despite QI-scrambling’s critical role in characterizing operational quantum systems, such as quantum computers, its exact quantification persists as a formidable challenge \cite{fauseweh2024quantum}. Theoretical frameworks, notably conformal field theories, and holographic models, have provided foundational insights \cite{asplund2015entanglement}, yet their relevance to experimental contexts remains constrained. To address this issue, researchers are creating ways to measure QI-scrambling and how it affects the use of quantum resources. Metrics such as tripartite information, which map information flux, elucidate the dynamics of scrambled data propagation \cite{PhysRevB.100.224302}. Similarly, out-of-time-order correlators, measuring non-chronological signal correlations, enable scrutiny of information delocalization \cite{PhysRevB.98.220303, PhysRevB.98.144304}. Operator entanglement further quantifies informational integration within systems, acting as a proxy for scrambling intensity \cite{PhysRevLett.129.050602}. Advances in simulating intricate quantum architectures, coupled with these tools, hold substantial promise for unraveling scrambling in realistic environments.

QI-scrambling has gained prominence as a pivotal research domain in quantum physics, with profound ramifications across theoretical and applied disciplines \cite{landsman2019verified}. Characterized by the non-local dispersal and entropic concealment of initial quantum states, scrambling impedes information retrieval via localized measurements, signifying its critical role in quantum state dynamics \cite{PhysRevA.94.040302}. This phenomenon is indispensable for elucidating the behavior of intricate quantum architectures, spanning black hole thermodynamics \cite{hayden2007black,daniel2025optimally}, many-body entanglement propagation \cite{PhysRevA.97.042330,PhysRevResearch.4.023095,PhysRevB.111.014439}, and information-theoretic protocols \cite{yunger2019entropic,sharma2021quantum}. Empirical studies utilizing trapped ions \cite{PhysRevLett.124.240505}, superconducting qubits \cite{mi2021information}, and Rydberg atom lattices \cite{PhysRevLett.126.200603} have validated the exponential delocalization of quantum information. Such results corroborate quantum chaos, mirroring classical chaotic analogs, where infinitesimal perturbations trigger systemic divergence. Recent theoretical advancements propose a universal framework for QI-scrambling in open systems \cite{PhysRevLett.131.160402}, resolving ambiguities in interpreting Loschmidt echo experiments and enhancing predictive capacity for quantum decoherence phenomena.  The effects of additional factors such as noise and errors on scrambling in open systems are not well understood. Exact numerical simulations for a qubit system coupled to an environment is provided \cite{PhysRevB.98.184416}.

 The examination of spin-star configurations considered in our case holds great promise for developments in quantum technologies \cite{joshi2022experimental}. Investigating the dynamics and associated correlations within these quantum many-body systems provide deeper understanding and insights into the behavioral aspects of such systems, allowing us to develop advanced quantum simulators and the realization of realistic quantum computing \cite{estarellas2018spin}. Additionally, the spin-star complex system exhibit unique properties which may lead to improvements in quantum information processing, therefore, enhancing the detection and measurement with high precision \cite{haddadi2019thermal,gassab2024geometrical}. 
Furthermore, examination of the quantum phase transitions and critical phenomena in the considered spin-star models could bring new quantum phases and emergent behavioral aspects in quantum communication and information protocols \cite{jakab2021quantum,haddadi2021suppressing,zahia2025explicit}.

 Consider a quantum many-body system wherein two local operators,  $\hat{W}_\mu^i(0)$ and $\hat{V}_\nu^j(0)$,  localized at lattice sites $i$ and $j$, respectively, indexed by $\mu$ and $\nu$. At the initial time $t=0$, these operators are commute, i.e.  $[\hat{W}_\mu^i(0), \hat{V}_\nu^j(0)]=0$. To investigate the influence of information scrambling on the local operator $\hat{W}_\mu^i(0)$, we shall examine its temporal evolution within the Heisenberg picture \cite{PhysRevX.9.031048}. The time evolution of the system is governed by the unitary operator $\hat{U}(t)$, which is subject to the many-body Hamiltonian $\hat{H}$. This relationship can be expressed as  $\hat{W}_\mu^i(t)= \hat{U}^\dagger(t) \hat{W}_\mu^i(0) \hat{U}(t) $, with the unitary constraint $i\hbar \partial_t \hat{U}(t)= \hat{H} \hat{U}(t)$ and $[\hat{H},\hat{W}_\mu^i(0)]\neq0$. A quantitative assessment of information scrambling can be achieved by using the commutator between the local operator  $\hat{V}_\nu^j(0)$ and its time-evolved counterpart $\hat{W}_\mu^i(0)$. This quantity is mathematically represented as \cite{PhysRevA.94.040302,PhysRevLett.122.040404}
 \begin{equation} \label{sc}
		\mathcal{C}_{\mu \nu}^{ij}(t) = \big\| \big[\hat{W}_\mu^i(t), \hat{V}_\nu^j(0)\big]\big\|^2_2.
\end{equation}
Here, $ \| \hat{O}\|_2^2= \text{Tr}[\hat{\rho}  \hat{O}^\dagger \hat{O}]$ being the square of the Hilbert-Schmidt norm of $\hat{O}$, where $\hat{\rho}$ can be defined either as the Gibb's density matrix,  
 $\hat{\rho}= e^ {-\beta \hat{H} }/\text{Tr}[ e^ {-\beta \hat{H} }]$, with $\beta$ is the inverse temperature,
or as the initial pure state $\hat{\rho}=| \psi \rangle \langle \psi |$.  In the context of a nonintegrable Hamiltonian, persistent $[\hat{W}_\mu^i(t),\hat{V}_\nu^j(0)]\neq 0$, manifested as a non-vanishing quantity $\mathcal{C}_{\mu \nu}^{ij}(t)$, i.e. $\mathcal{C}_{\mu \nu}^{ij}(t)>0$, engenders a significant and sustained decrease in the negative parts of the Eq.  (\ref{sc}) \cite{PhysRevLett.122.040404}. Conversely, within the framework of a non-scrambling Hamiltonian, the negative parts of $\mathcal{C}_{\mu \nu}^{ij}(t)$  exhibit a revival, approaching a value close to unity, accompanied by a near-commuting relationship between $\hat{W}_\mu^{i}(t)$ and $\hat{V}_\nu^{j}(0)$, i.e. $[\hat{W}_\mu^{i}(t),\hat{V}_\nu^{j}(0)]\cong 0$ \cite{PhysRevResearch.4.023095}.

 The motivation underpinning this paper is multifaceted. Primarily, we seek to redefine the relationship between QI-scrambling and its boundaries within the Maligranda inequality. The modified version of this inequality is used before to identify the upper limits of the uncertainty relation between two operators, but using weak limits\cite{PhysRevA.95.052117}. This paper aims to determine the extent to which these limits converge and to identify the potential factors contributing to any discrepancies in their behavior. Furthermore, the maximum boundary of QI-scrambling for general unitary-Hermitian or arbitrary Hermitian operators is determined. We examine the nature of QI-scrambling within a spin-star model, considering local operators, both unitary-Hermitian single-spin and multi-spin (arbitrary Hermitian) operators. Further, the feasibility of controlling the scrambling within the system is discussed through two different initial states.

\section{Scrambling limits by Maligranda inequality}
If $x$ and $y$ are nonzero vectors in a normed linear space, the Maligranda inequality is given by \cite{maligranda2006simpl}
\begin{equation} \label{e22}
	\frac{\|x - y\| - |\|x\| - \|y\||}{\min\{\|x\|, \|y\|\}} \leq\left\| \frac{x}{\|x\|} - \frac{y}{\|y\|} \right\| \leq \frac{\|x - y\| + |\|x\| - \|y\||}{\max\{\|x\|, \|y\|\}},
\end{equation}
where $x$ and $y$ are nonzero vectors, and non-negativity property of the norm with $\|x + y\| \leq \|x\| + \|y\|. $

Let, $|x \rangle= \bar{A} |\bar{x}\rangle$ and $  |y\rangle= \bar{B} |\bar{y}\rangle $, where $ \bar{A} = \hat{A}-\langle \hat{A}\rangle $ and $ \bar{B} = \hat{B}-\langle \hat{B}\rangle $, with $\hat{A}, \hat{B}$ are any arbitrary operators. Then, we have
\begin{eqnarray} \label{e4}
	\|x\| &=&  \Delta \hat{A}, \  \|y\| =  \Delta \hat{B},\ \text{and}, \|x-y\| =  \Delta (\hat{A}-\hat{B}),   
\end{eqnarray}
where, $\Delta^2 \hat{O}=\langle \hat{O}^2 \rangle-\langle \hat{O} \rangle^2 $ is the variance of operator $ \hat{O}$.  Also, we can get
\begin{equation} \label{e5}
	\left\| \frac{x}{\|x \|} - \frac{y}{\|y\|} \right\|
	=  \left[ 2 - \frac{ 2\ \text{Re}\big(\text{Cov} (\hat{A}, \hat{B})\big)}{\Delta \hat{A} \Delta \hat{B}} \right]^{\frac{1}{2}}.
\end{equation}
Here, $\text{Cov}(\hat{A}, \hat{B})= \langle \hat{A}^\dagger \hat{B} \rangle - \langle \hat{A}^\dagger \rangle \langle \hat{B} \rangle$ is the covariance. From Eqs. (\ref{e22}) to (\ref{e5}), we get
\begin{eqnarray} \label{e6}
	2- \bigg(\frac{\Delta (\hat{A} - \hat{B}) - |\Delta \hat{A} - \Delta \hat{B}|}{ \min[\Delta \hat{A}, \Delta \hat{B} ]}\bigg)^2  & \geq &  \frac{2\text{Re}[\text{Cov}(\hat{A}, \hat{B}))]}{\Delta \hat{A} \Delta \hat{B}} \nonumber \\   & \geq & 2-  \bigg( \frac{ \Delta (\hat{A} - \hat{B}) + |\Delta \hat{A} - \Delta \hat{B}|}{\max [\Delta \hat{A}, \Delta \hat{B} ]}\bigg)^2.
\end{eqnarray}

If we choice $ \hat{V}_\nu^{j}(0) \hat{W}_\mu^{i}(t)=\hat{A}$ and $  \hat{W}_\mu^{i}(t) \hat{V}_\nu^{j}(0)=\hat{B}$, with Eq. (\ref{sc}), we get
\begin{equation} \label{sc2}
		\mathcal{C}_{\mu \nu}^{ij}(t) = \langle \hat{A}^\dagger \hat{A} \rangle + \langle \hat{B}^\dagger \hat{B} \rangle- \langle \hat{A}^\dagger \hat{B} \rangle-\langle \hat{B}^\dagger \hat{A} \rangle.
\end{equation}
Using Eqs. (\ref{e6}) and (\ref{sc2}), we get the QI-scrambling bounds as

\begin{eqnarray} \label{cc}
	\mathcal{L}_{\mu \nu}^{ij}(t) & \leq & \mathcal{C}_{\mu \nu}^{ij}(t)  \leq \mathcal{U}_{\mu \nu}^{ij}(t).
\end{eqnarray}
Here
\begin{eqnarray}
		&\mathcal{L}_{\mu \nu}^{ij}(t)= \Delta \hat{A} \Delta \hat{B} ( \mathcal{J}^2-2) + \| \hat{A}\|_2^2 + \| \hat{B}\|_2^2- 2 \text{Re}[\langle \hat{B}^\dagger \rangle \langle \hat{A} \rangle]\ ,  \nonumber \\
		&\mathcal{U}_{\mu \nu}^{ij}(t)= \Delta A \Delta B ( \mathcal{K}^2-2)+  \| \hat{A}\|_2^2 + \| \hat{B}\|_2^2- 2 \text{Re}[\langle \hat{B}^\dagger \rangle \langle \hat{A} \rangle]\ ,
\label{eq10}
\end{eqnarray}
with 
\begin{eqnarray}\label{hh}
		&\mathcal{J}=\frac{\Delta (\hat{A} - \hat{B}) - |\Delta \hat{A} - \Delta \hat{B}|}{ \min[\Delta \hat{A}, \Delta \hat{B} ]}\ , \quad
		\mathcal{K}=  \frac{ \Delta (\hat{A} - \hat{B}) + |\Delta \hat{A} - \Delta \hat{B}|}{\max [\Delta \hat{A}, \Delta \hat{B} ]}\ .
\end{eqnarray}

If we choice the operators $\hat{W}_\mu^{i}(t), $ and $ \hat{V}_\nu^{j}(0)$ are unitary and Hermitian, then
\begin{eqnarray}
		&\Delta \hat{A} = \sqrt{1 - \langle \hat{W}_\mu^{i}(t) \hat{V}_\nu^{j}(0) \rangle \langle\hat{V}_\nu^{j}(0) \hat{W}_\nu^{i}(0) \rangle}, \nonumber\\
		& 
		\Delta \hat{B} = \sqrt{1 - \langle \hat{V}_\nu^{j}(0) \hat{W}_\mu^{i}(t)  \rangle \langle \hat{W}_\mu^{i}(t) \hat{V}_\nu^{i}(0) \rangle}.
\end{eqnarray}
Thus
\begin{equation}
	\mathcal{J}= \mathcal{K}=\frac{\Delta (\hat{A} - \hat{B})}{\Delta \hat{A}}.
\end{equation}
Consequently, the upper and lower bounds coincide with QI-scrambling as
\begin{equation}\label{MS}
	\mathcal{L}_{\mu \nu}^{ij}(t)=\mathcal{U}_{\mu \nu}^{ij}(t)=\mathcal{C}_{\mu \nu}^{ij}(t) = 2(1-\text{Re}[\langle \hat{A}^2 \rangle]).
\end{equation}

However, for general local \textit{Hermitian} operators, the situation is more complex. Here, the equivalence of the upper and lower bounds—that is, whether they converge to the same value—is not guaranteed and depends on both the choice of local operators and the system’s initial state.  This dependence arises because the QI-scrambling measure, which quantifies information spreading in the system, is sensitive to how the initial state interacts with the operator dynamics.

\begin{figure}[!h]
	\centering
	\includegraphics[width=0.42\textwidth, height=5cm]{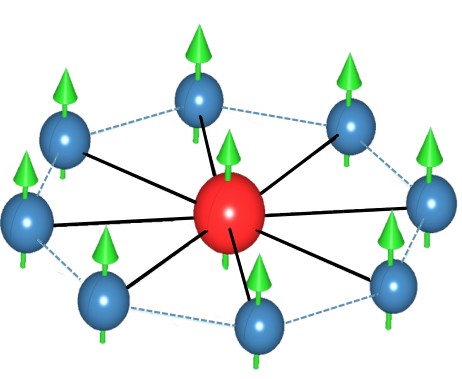}
	\caption{A schematic representation of a central qubit $\hat{S}^a$ (red) surrounded by a collection of peripheral qubits $\hat{S}^i$ (blue). The central qubit interacts with each peripheral qubit via a coupling constant $J_1$, while the peripheral qubits interact with each other via a coupling constant $J_2$.}
	\label{f1}
\end{figure}

To elucidate the limits of QI-scrambling, we propose a spin-star architecture consisting of a single two-level ancilla coupled to $N$ identical qubits (depicted in Fig.\ref{f1}). The ancilla interacts isotropically with all peripheral qubits via $XXX$ coupling with a coupling strength $J_1$. Furthermore, the peripheral qubits provide mutual interactions with the nearest neighbors, with the coupling strength $J_2$, while having periodic boundary conditions. The Hamiltonian of the system consists of the free systems and interaction of the ancilla and the outer qubits, as well as the interactions among the outer qubits themselves. The physical Hamiltonian governing this system can be expressed as ($\hbar=1$) \cite{arisoy2021few}
\begin{equation}
	\hat{H}= \hat{H}_{free}+ \hat{H}_{an-sp}+ \hat{H}_{sp-sp}\ .
\end{equation}
Here, the free Hamiltonian $\hat{H}_{free}= \omega_a \hat{S}_{z}^a+\omega_s\sum_{i=1}^{N}\hat{S}_{z}^i$, where $\omega_a$ and  $\omega_s$ represent the frequencies of the ancilla and the outer qubits system, respectively. The operators $\hat{S}_{z}^a$ and $\hat{S}_{z}^i$ correspond to the $z$-Pauli matrices for the ancilla and $i^{th}$ outer qubits. The interaction Hamiltonian between the ancilla and the outer qubits is represented by $ \hat{H}_{an-sp}=J_1\sum_{i=1}^{N}\big( \hat{S}_{x}^a\hat{S}_{x}^i+\hat{S}_{y}^a\hat{S}_{y}^i+\hat{S}_{z}^a\hat{S}_{z}^i \big)$. Additionally, the qubit-qubit interaction Hamiltonian is expressed as  $\hat{H}_{sp-sp}= J_2\sum_{i=1}^{N} \vec{S}^i.\vec{S}^{i+1}$ with the coupling $J_2 $ and periodic boundary condition is imposed such that $\vec{S}^{N+1}= \vec{S}^{1} $.

This work investigates the correlation between the upper bounds $\mathcal{U}_{\mu \nu}^{ij}(t)$, lower bounds $\mathcal{L}_{\mu \nu}^{ij}(t)$, and QI-scrambling $\mathcal{C}_{\mu \nu}^{ij}( t)$. To this end, understanding the interplay of these metrics is crucial for unraveling the dynamics of quantum information propagation within a system. These limits alignment reveals whether $\mathcal{C}_{\mu \nu}^{ij}(t)$ is accurately capturing the system's scrambling dynamics and can potentially identify phase transitions or changes in system behavior. A significant divergence between $\mathcal{C}_{\mu \nu}^{ij}(t)$ and the bounds may indicate a transition from ordered to chaotic states.

Moreover, we postulate that the internal ancilla qubit resonates with the outer qubits ensemble, $\omega_a=\omega_s=\omega_0$. The outer qubits are coupled to the ancilla via a coupling constant $J_1=1$ and with each other $J_2=0.5$. This analysis will encompass both the thermal state and the pure state. For the initial pure state, we assume the ancilla in excited and the outer qubits in the ground states, with $|\psi \rangle=|\uparrow \rangle_a \otimes |\downarrow\rangle^ {\otimes N} $. 

\section{The maximum limit of scrambling.}
 \quad The scrambling phenomenon is not exclusively contingent upon the dimensionality of local operators, a concept frequently highlighted in certain studies. Nevertheless, in the present analysis, irrespective of the dimensions of local operators, the maximum value of QI-scrambling is constrained to a value less than or equal to four, provided these operators retain their Hermitian-unitary properties. For this proposition, QI-scrambling using unitary-Hermitian operators $\hat{W}_{\mu}^{i}(0), \hat{V}_{ \nu}^{j}(0) $ is expressed as
 \begin{eqnarray}\label{US}
 	\mathcal{C}_{\mu \nu}^{ij}(t)& = \left\|\hat{W}_{\mu}^{i}(t) \hat{V}_{ \nu}^{j}(0) - \hat{V}_{ \nu}^{j}(0) \hat{W}_{\mu}^{i}(t) \right\|^2_2 \nonumber \\&
 	\leq \bigg(\left\| \hat{W}_{\mu}^{i}(t) \hat{V}_{ \nu}^{j}(0) \right\|_2 + \left\|\hat{V}_{ \nu}^{j}(0) \hat{W}_{\mu}^{i}(t) \right\|_2\bigg)^2
 	=4
 \end{eqnarray}
 
 Assumed the two local operators posited that the ancilla remains in a stationary state while at least one of the outer qubits undergoes temporal evolution, as represented by
 \begin{equation*}
 	\hat{W}_\mu^{i}(0)=\{\hat{S}_\mu^i \}_{\mu=x,z}^{i=1,2,\dots} \quad \text{and} \ \hat{V}_\nu^{j}(0)=\{\hat{S}_\nu^a\}_{\nu=x,z}\ .
\end{equation*}
\begin{figure}
	\centering
	\includegraphics[width=0.99\textwidth, height=9cm]{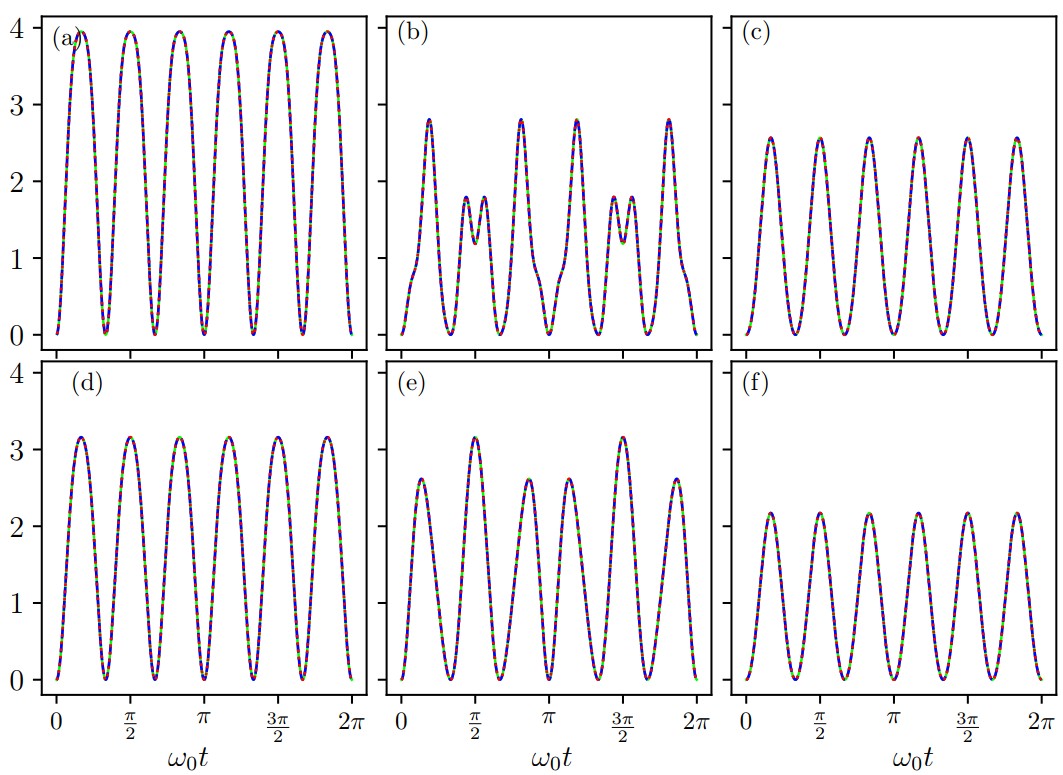}
		\vspace{-0.2cm}
	\caption{The QI-scrambling $\mathcal{C}_{\mu \nu}^{ij}(t)$ (blue curve), lower bounds $\mathcal{L}_{\mu\nu}^{ij}(t)$ (red-curve), and upper bounds $\mathcal{U}_{\mu\nu}^{ij}(t)$ (green curve), where $N=2$, $J_1=1$, $J_2=0.5$, and the lattice sites $i=1$, and $j=a$. The initial state is the pure state $|\psi\rangle=|\uparrow \rangle_a \otimes |\downarrow\rangle^ {\otimes 2} $. (a) $\mu=\nu=z$ (b) $\mu=\nu=x$. (c), and (d) are the same as (a), and (b) respectively, with the initial thermal state at $\beta=10$. }
	\label{f2}
\end{figure}	
  \quad Figure \ref{f2} displays the scrambling limits $\mathcal{L}_{\mu\nu}^{ij}(t)$, $\mathcal{C}_{\mu \nu}^{ij}(t)$, and $\mathcal{U}_{\mu\nu}^{ij}(t)$ for $N=2$, considering both pure and thermal initial states at inverse temperature $\beta=10$.  It is observed that the maximum attainable scrambling, regardless of the choices $\hat{W}_\mu^{i}(0)$ and $\hat{V}_\nu^{a}(0)$, does not surpass the theoretical upper bound delineated by Eq.(\ref{US}). A noteworthy finding is the complete congruence of the three functions by Eq.(\ref{MS}), a consequence of the initial operators' unitary-Hermitian nature. Moreover, QI-scrambling in the pure state outperforms its thermal counterpart for $\hat{W}_z^{1}(0)$, $\hat{V}_z^{a}(0)$ (Fig.\ref{f2}(a)). In contrast, the thermal state exhibits a similar periodic pattern with a diminished amplitude (Fig.\ref{f2}(d)). For the specific choice of  $\hat{W}_x^{1}(0)$, $\hat{V}_x^{a}(0)$, QI-scrambling oscillates periodically, adhering to the relation $\omega_0t = n\pi ,n=\{0,1,\dots\}$ but fails to reach its maximum value. The circular symmetry of the system induces a periodic evolution of its quantum state. This periodicity is the primary factor influencing the scrambling behavior. The periodic behavior may be attributed to the limited number of qubits, which ensures a periodic distribution of information. However, it is noticeable that the system's growing dimensionality introduces more randomness in comparison. It is also noteworthy that the perfect alignment between the upper and lower scrambling bounds persists as long as the local operators represent a single particle in a unitary-Hermitian state, irrespective of changes in the number of qubits, system parameters, or dimensions.

\section{Multi-partite Local Operators (Block Operators).} 
Let the operators $\hat{W}_\mu^{i}(0)$ and $\hat{V}_\nu^{j}(0)$ consist of $m$ and $n$ number of multi-partite system. Such that
\begin{eqnarray*}
		\hat{W}_\mu^{i}(0)= \sum_{i=1}^{m} \hat{w}_\mu^{i}(0)\ ,\quad
		\hat{V}_\nu^{j}(0) = \sum_{j=1}^{n} \hat{v}_\nu^{j}(0)\ .
\end{eqnarray*}
Here, $\hat{w}_\mu^{i}(0)$, and $\hat{v}_\nu^{j}(0)$ are arbitrary local operators. Then, the upper bound of QI-scrambling is 
\begin{eqnarray}
		\mathcal{C}_{\mu,\nu}^{ij}(t) &= \left\| \left[ \sum_{i}^{m} \hat{w}_\mu^{i}(t), \sum_{j}^{n} \hat{v}_\nu^{j}(0) \right] \right\|_2^2 \nonumber \\&
		\leq\sum_{l}^{i}\sum_{j}^{n}\left\|  [\hat{w}_\mu^{i}(t), \hat{v}_\nu^{j}(0)] \right\|_2^2 \nonumber \\&= n^2 m^2 \mathcal{M}\ ,
		\label{smm}
\end{eqnarray}

where $\mathcal{M}=\max [\mathcal{C}_{\mu\nu}^{ij}(t)]$ is the maximum scrambling $ \forall\quad \{i,j\} $.

Consider the scenario wherein the operator $\hat{W}_{\mu}^{i}(0)$ comprises $m$ outer qubit, while $\hat{V}_{\nu}^{j}(0)$ is associated with the ancilla. In this scenario, the central ancilla maintains a constant state while a cohort of outer qubits undergoes temporal evolution, expressed as:
	\begin{eqnarray*}
			\hat{W}_\mu^{i}(0)=\bigg\{\sum_{i}^{m}\hat{S}_\mu^i \bigg\}_{\mu=x,z} , \ \text{and} \ \hat{V}_\nu^{a}(0)=\{\hat{S}_\nu^a\}_{\nu=x,z}\ .
	\end{eqnarray*}
	
\begin{figure}[!h]
\centering
\includegraphics[width=0.99\textwidth, height=9cm]{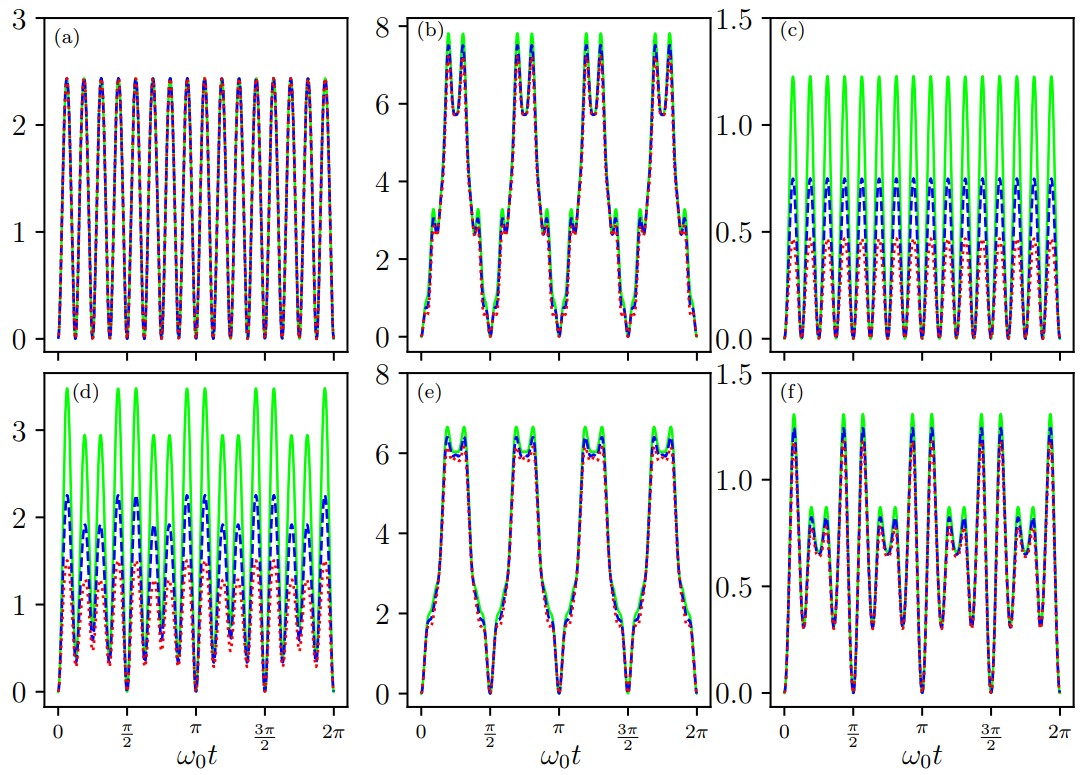}
\vspace{-0.5cm}
	\vspace{-0.1cm}
\caption{The QI-scrambling $\mathcal{C}_{\mu \nu}^{ij}(t)$ (blue curve), lower bounds $\mathcal{L}_{\mu\nu}^{ij}(t)$ (red-curve), and upper bounds $\mathcal{U}_{\mu\nu}^{ij}(t)$ (green curve), where $N=7$, $J_1=1$, $J_2=0.5$, and the lattice sites $i$ ranges from $1$ to $3$, and $j=a$. The initial state is the pure state $|\psi\rangle=|\uparrow \rangle_a \otimes |\downarrow\rangle^ {\otimes 7} $. (a) $\mu=\nu=z$  (b)  $\mu=\nu=x$, (c) $\mu=x, $ and $\nu=z$.  (d), (e), and (f) are the same as (a), (b), and (d) respectively, with the initial thermal state at $\beta=10$.}
	\label{f3}
\end{figure}	

A clear distinction can be made between the scrambling limits $\mathcal{L}_{\mu\nu}^{ij}(t)$, $\mathcal{C}_{\mu \nu}^{ij}(t)$, and $\mathcal{U}_{\mu\nu}^{ij}(t)$ as seen in Fig.\ref{f3} for both pure and thermal states. This discrepancy is contingent upon the selection of the initial state for the local operators. In the case where $\hat{W}_x^{i}(0)=\sum_{i=1}^{3}\hat{S}_x^{i}$, and $\hat{V}_x^{a}(0)$, the inequality provided in Eq.(\ref{cc}) holds, with the discrepancy arising due to the distinction between $\mathcal{J}$ and $\mathcal{K}$ as defined in Eq.(\ref{hh}). This disparity originates from the inherent differences between $\mathcal{J}$ and $\mathcal{K}$. In the instance of $\hat{W}_z^{i}(0)=\sum_{i=1}^{3}\hat{S}_z^{i}$, and $\hat{V}_z^{a}(0)$, a divergence is observed between the $\mathcal{L}_{\mu\nu}^{ij}(t)$, and $\mathcal{U}_{\mu\nu}^{ij}(t)$, accompanied by scrambling in the thermal state. In contrast, complete convergence of the three bounds is evident in the pure state where $\mathcal{J}=\mathcal{K}$. Notwithstanding the apparent behavioral similarity among $\mathcal{U}_{\mu\nu}^{ij}(t)$, $\mathcal{C}_{\mu\nu}^{ij}(t)$, and $\mathcal{L}_{\mu\nu}^{ij}(t)$ in the thermal regime, they exhibit non-identical characteristics. The gap between the three bounds becomes larger in the pure state because of the significant difference between $\Delta \hat{A}$ and $\Delta \hat{B}$. This difference grows as the number of outer qubits increases, raising the maximum values of all three functions. Despite the substantial divergence between the three bounds, it remains confined within the maximum scrambling range established by Eq.(\ref{smm}). Similarly, alignment occurs when $ N = 7 $, where $\hat{W}_z^{i}(0)=\sum_{i=1}^{4}\hat{S}_z^i$, and $\hat{V}_x^{j}(0)=\hat{S}_x^{a}$, with the initial state being thermal.  However, in other scenarios where these conditions are not met, the scrambling behavior deviates, with a clear divergence between the upper and lower bounds.

A noteworthy observation is that the scrambling behavior remains periodic over time even though one of the local operators is not strictly unitary. This periodic behavior can be clarified by the unitary-Hermitian nature of the local operator $\hat{V}_\nu^a(0)$, which helps preserve the periodic scrambling structure. Also, the periodicity may result from selecting consecutive outer spins to commute with the central ancilla, thereby imposing a fixed distance irrespective of the number of outer spins. Furthermore, the results indicate that the measurement of quantum information scrambling reduces as the number of outer spins increases. This behavior suggests that a more extensive system size mitigates the scrambling effect, likely due to the increased dimensionality of the Hilbert space, which spreads the information over a broader configuration space, thereby reducing localized scrambling. It is incontrovertible that varying the initial state yields different outcomes, but this depends primarily on the choice of local operators and their number.

\begin{figure}[!h]
	\centering
	\includegraphics[width=0.99\textwidth, height=9cm]{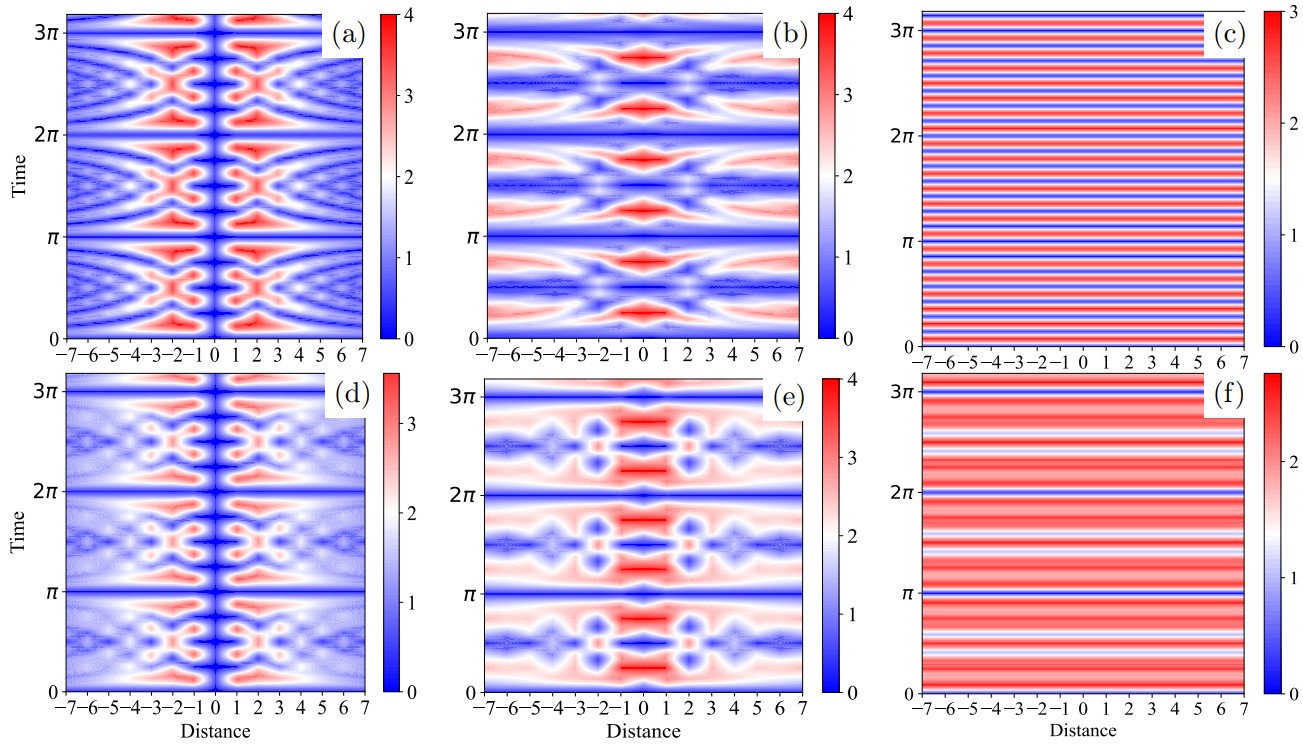}
	\vspace{-0.3cm}
	\caption{Contour plot of QI-scrambling over time and distance for  $J_1=1$, and $J_2=0.5$. The upper panels: the initial pure state $|\psi\rangle=|\uparrow \rangle_a \otimes |\downarrow\rangle^ {\otimes 6} $ for (a) $\mathcal{C}_{zz}^{1a}(t)$, (b) $\mathcal{C}_{xx}^{1a}(t)$, (c) $\mathcal{C}_{zz}^{ka}(t)$ where $k$ ranges from $1$ to $3$. Lower panels: the initial thermal state with $\beta=1$ and the same parameters as in the upper panels.}
	\label{c1}
\end{figure}
The contour plot of QI-scrambling under three distinct initial local operators $\mathcal{C}_{zz}^{1a}(t)$, $\mathcal{C}_{xx}^{1a}(t)$ or $\mathcal{C}_{zz}^{ka}(t)$, $k$ tacks the summation from $1$ to $3$, for both pure and thermal states is depicted in Fig.\ref{c1}. A notable observation is the periodic temporal evolution of scrambling under the three scenarios in the spin-star model. The individual local operators for both initial states show that as the distance (number of particles) increases, the maximum bounds of scrambling diminish, with overlapping propagation waves occurring at every interval of $\pi$. The oscillatory behaviour of $\mathcal{C}_{zz}^{1a}(t)$ and $\mathcal{C}_{xx}^{1a}(t)$ indicate that the system is neither scrambling nor chaotic. This behavior is because, even in the larger system considered by $N=7$, the chosen initial state is fully polarized except for the central spin, resulting in a minimal Hilbert space dimension. In this system, one can probably even calculate the squared commutator analytically. The block-local operators exhibit distance-independent scrambling. This behavior arises from the block-local operator $\sum_{i=1}^{3}\hat{S}_z^i$, which inherently selects three consecutive qubits, thereby imposing a fixed distance irrespective of the number of outer qubits. Overall, the maximum bounds of QI-scrambling are more pronounced in the initial pure state than in the initial thermal state.

Within the context of examining scrambling on the outer spins while excluding the ancilla from the local operators, the local operators consist of some spins from the outer spins as
\begin{equation*}
		\hat{W}_\mu^i(0)=\bigg\{\sum_{i=1}^{m}\hat{S}_\mu^i\bigg\}_{\mu=x,z}, \ \text{and} \ \hat{V}_\nu^j(0)=\bigg\{\sum_{j=m+1}^{N}\hat{S}_\nu^j\bigg\}_{\nu=x,z}.
\end{equation*}

\begin{figure}[!h]
\includegraphics[width=0.99\textwidth, height=9cm]{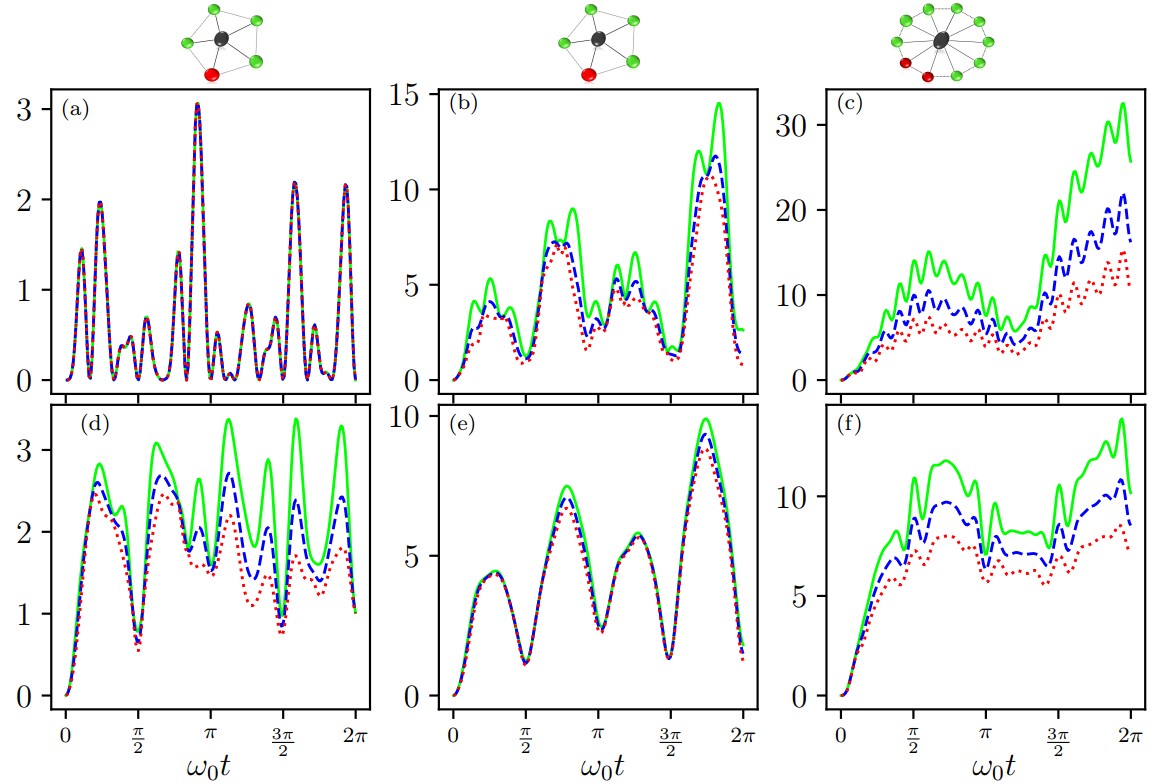}
\vspace{-0.3cm}
\caption{The QI-scrambling $\mathcal{C}_{\mu \nu}^{ij}(t)$ (blue curve), lower bounds $\mathcal{L}_{\mu\nu}^{ij}(t)$ (red-curve), and upper bounds $\mathcal{U}_{\mu\nu}^{ij}(t)$ (green curve), where $J_1=1$, $J_2=0.5$. The initial state in the pure state $|\psi\rangle=|\uparrow \rangle_a \otimes |\downarrow\rangle^ {\otimes N} $. (a) $N=5$, $ \hat{V}_z^{j}(0)= \hat{S}_z^1 $ and $\hat{W}_z^{i}(0)=\sum_{i=2}^{5}\hat{S}_z^i$.  (b)  $N=5$, $ \hat{V}_x^j(0)=\hat{S}_x^1 $ and $\hat{W}_x^{i}(0)=\sum_{i=2}^{5}\hat{S}_x^i$. (c) $N=10$, $\hat{V}_x^j(0)=\sum_{j=1}^{2}\hat{S}_x^j$ and $\hat{W}_z^i(0)=\sum_{k=3}^{10}\hat{S}_x^i$. (d), (e) and (f) are the same as (a), and (b) respectively, with the initial thermal state at $\beta=1$.}
	\label{f4}
\end{figure}

 As depicted in Fig. \ref{f4}, a notable divergence between the three bounds, yet maintains a consistent qualitative trend, suggesting the efficacy of the functions $\mathcal{U}_{\mu\nu}^{ij}(t)$ and $\mathcal{L}_{\mu\nu}^{ij}(t)$ in quantifying scrambling. A general correlation emerges between the number of spins within $\hat{W}_\mu^{i}(0)$ and the disparity between the three bounds. A marked convergence is observed among the upper bounds, scrambling, and lower bounds within the pure state, with the complete overlap of all three metrics, when $ \hat{V}_z^{i}(0)=\hat{S}_z^1 $ and $\hat{W}_z ^i(0)=\sum_{i=2}^{5}\hat{S}_z^i$. Regardless of the unitary-Hermitian nature of $\hat{V}_\mu^j(0)$ (as illustrated in Figs \ref{f4} (a), (b), (d), (e)), the scrambling behavior transitions to a stochastic pattern. However, the peak of scrambling behavior increases with time growing for $ \hat{V}_x^j(0)=\hat{S}_x^1 $ and
 $\hat{W}_x^{i}(0)=\sum_{i=2}^{5}\hat{S}_x^i$ because the total $ \hat{S}_x$ is not conserved versus $S_z$. The chosen initial state is fully polarized except for the central spin. The initial thermal state results in a diminished maximum scrambling value compared with the pure state. The lower values exhibit a more pronounced increase and fail to converge to zero over time, indicating an increase in system chaos. Increasing the outer spins increases the maximum scrambling with time (both local operators are Hermitian Figs \ref{f4} (c), (f))). This behavior implies that outer spins are the primary drivers of scrambling within the system. A reduction in the number of outer spins is observed to correlate with a decrease in scrambling, where the scrambling dynamics take in a small Hilbert space. In general, scrambling does not attain the maximum value predicted by Eq. (\ref{smm}) but remains consistently lower. Notably, this condition exhibits the most chaotic scrambling behavior among all investigated states.

\begin{figure}[!h]
\centering
\includegraphics[width=0.99\textwidth, height=9cm]{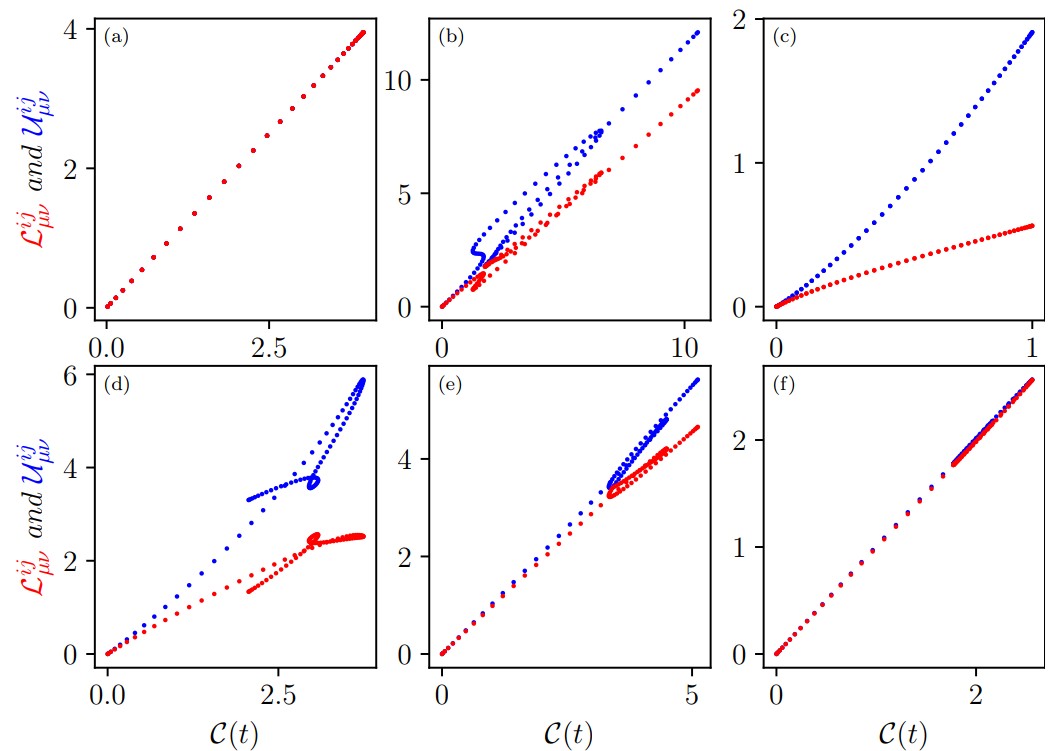}
\vspace{-0.3cm} \caption{Scattered plot between upper bound $\mathcal{U}_{\mu\nu}^{ij}(t)$ (blue dots), and $\mathcal{L}_{\mu\nu}^{ij}(t)$ (red dots) as functions of $\mathcal{C}_{\mu\nu}^{ij}$ for $\omega_0 t=\pi/4$ $J_1=1$, and $J_2=0.5$. The initial state is the pure state $|\psi\rangle=|\uparrow \rangle_a \otimes |\downarrow\rangle^ {\otimes N} $. (a) $N=5$,  $\hat{W}_z^{i}(0)=\sum_{i=1}^{4}\hat{S}_z^{i}$, $\hat{V}_z^{j}(0)=\hat{S}_z^{a}$. (b)  $N=5$, $\hat{W}_x^{i}(0)=\sum_{i=1}^{4}\hat{S}_x^{i}$, $\hat{V}_x^{j}(0)=\hat{S}_x^{a}$. (c) $N=7$, $\hat{W}_z^{i}(0)=\sum_{i=1}^{4}\hat{S}_z^i$, $\hat{V}_x^{j}(0)=\hat{S}_x^{a}$. (d), (e) and (f) are the same as (a), and (b) respectively, with the initial  thermal state at $\beta=1$.}
\label{D1}
\end{figure}

In Fig. \ref{D1}, we present comparing the lower bounds $\mathcal{L}_{\mu\nu}^{ij}(t) $ (red dots), upper bounds $\mathcal{U}_{\mu\nu}^{ij}(t) $ (blue dots), and QI-scrambling $\mathcal{C}_{\mu\nu}^{ij}(t) $ for various cases of the spin-star model with a central ancilla interacting with an ensemble of $N $ qubit particles. The parameters for the plots are set to $\omega_0 t = \pi/4 $, $J_1 = 1 $, and $J_2 = 0.5 $, with variations in the initial states and operators used for $\hat{W}_\mu^i(0) $ and $\hat{V}_\nu^j(0) $. Figure \ref{D1} (a)  shows a case where the system is assumed as a pure state $|\psi\rangle = |\uparrow\rangle_a \otimes |\psi\rangle^{\otimes 5} $, with $\hat{W}_z^i(0) = \sum_{i=2}^4 \hat{S}_z^i $ and $\hat{V}_z^j(0) = \hat{S}_z^i $. The red and blue curves align closely, indicating a linear correlation between the upper and lower bounds as a function of QI-scrambling, suggesting tight bounds for QI-scrambling. Figure \ref{D1} (b) shows the same initial state but with $\hat{W}_\mu^i(0) = \sum_{i=2}^4 \hat{S}_z^i $ and $\hat{V}_\nu^j(0) = \hat{S}_z^a $. Despite the discrepancy between the lower and upper bounds of scrambling, the relationship between them remains linear within a certain range before transitioning to a nonlinear regime. Figure \ref{D1}(c) features the same system but with $N = 7 $ and a different choice of operators: $\hat{W}_\mu^i(0) = \sum_{i=2}^4 \hat{S}_z^i $ and $\hat{V}_\nu^j(0) = \hat{S}_x^a $. The plot shows that the lower bounds correlate more with scrambling than the upper bounds. By fitting the scattered data, linear regression analysis suggests that the relationship between the scrambling limits remains a simi-linear function. Figures \ref{D1} (d, e, f) repeat the first three scenarios but with the initial state prepared as a thermal state with inverse temperature $\beta = 1 $. In these cases, the discrepancy between the upper and lower bounds persists. At the same time, the fitting of scattered data more closely reflects a simi-linear relationship between QI-scrambling and its limits. As the local operator size increases or the operators vary, the gap between the bounds widens, suggesting more intricate scrambling dynamics. The thermal state exhibits smoother correlations between scrambling limits compared to the pure state but still demonstrates teeny distinct deviations between the bounds in specific configurations.

\section{Significance the Results} \label{sec:Si}
\vspace{-0.5cm}
\quad The significance of the results entails that the dependence of the scrambling bounds on the local operator selection highlights the necessity to pay attention to the operator's choice to describe the quantum information scrambling properly. Our results definitively establish that the maximum value of QI-scrambling remains invariant (at four) across system dimensions, provided that local operators adhere to unitary-Hermitian constraints. We have identified at least one specific instance involving two local unitary-Hermitian operators, where the maximum value of QI-scrambling remains constant regardless of the system's dimensionality. This observation indicates that maximum and minimum bounds of QI-scrambling can occur independently of system size. Here, we discuss QI-scrambling's maximum and minimum limits rather than its propagation characteristics. As our bounds have not been previously investigated, we employed Maligranda's inequality, a powerful tool for studying the upper and lower bounds of given operators \cite{PhysRevA.95.052117}. The propagation in the Spin-star model is constrained by small numbers of spins, rendering the investigation of propagation using Lieb–Robinson bound or Lyapunov exponent boundaries superfluous \cite{PhysRevLett.97.050401,cotler2018out}. Another critical aspect of our research is the case of non-unitary Hermitian operators. As demonstrated in Eq.(\ref{smm}), the maximum value of QI-scrambling can substantially exceed its actual value. Consequently, determining the greatest upper bound $\mathcal{U}_{\mu \nu}^{ij}(t)$ and the greatest lower bound $\mathcal{L}_{\mu \nu}^{ij}(t)$ is essential.
In simpler terms, the upper bounds of scrambling tell us that there is a maximum limit to how scrambling precisely we can know certain pairs of local operators. This means that not only is there a fundamental limit to how precisely we can measure these properties together, but there is also a limit to how imprecise these measurements can be. This duality provides a complete understanding of the limitations and behavior of scrambling systems. The fact that unitary-Hermitian operators yield invariant maximum scrambling bounds offers valuable insights into the design of quantum systems and protocols that operate within the optimal scrambling regime. The results indicated that the reduction in peak QI-scrambling intensity with increasing qubit count is specific to local unitary-Hermitian operators. This behavior arises because, as the number of qubits increases, the Hilbert space grows, diluting the influence of a single local operator across the system. In larger quantum systems, the effect of a local perturbation becomes less significant relative to the total degrees of freedom, resulting in a decrease in the maximum values of scrambling \cite{PhysRevA.97.042330,PhysRevA.100.032309}. For Hermitian operators, our results—illustrated in Figs \ref{f3} and \ref{f4}—demonstrated the opposite effect: QI-scrambling increases as the qubit count grows. Unlike unitary-Hermitian operators, general Hermitian operators lack the constraint of unitarity, allowing them to generate more complex dynamics. However, the maximum bounds of scrambling do not exceed the proved inequality (\ref{smm}).
Additionally, the scrambling bounds and stochastically due to the initial state can merge, creating more opportunities for better quantum state preparation and stabilization with prospects in quantum information processing in computation, communication, sensing, and simulation. By understanding how local operator selection, scrambling behavior, and multi-partite interactions affect a quantum system, researchers can better design quantum systems that are either robust against errors or deliberately scrambled to secure information. It implies that not only can we not prepare quantum states with arbitrarily small scrambling, but also we cannot prepare states with arbitrarily large scrambling. A fundamental limit exists to how precise a scrambling quantum state can be. These results inform laboratory studies, enabling the prediction of upper and lower bounds for scrambling, which might exhibit minor deviations from theoretical predictions. By comparing experimental data with theoretical predictions of scrambling limits, we can validate the accuracy of the models under investigation and refine them as needed.

\section{Conclusion}
 We investigated the QI-scrambling scheme uniformly constrained by or equivalent to the Maligranda bounds. Under the local operators, which are unitary-Hermitian, the maximum value of QI-scrambling is invariant regardless of system dimensions. This value indicates that QI-scrambling can occur independently of system dimensions. Our results highlight the critical influence of local operator selection on the alignment between the derived bounds and scrambling behavior. When the chosen local operator comprises unitary-Hermitian operators, perfect consistency is observed. Conversely, non-unitary local operators may yield valid correlations but lack consistency, representing the system's upper scrambling limit.
 Furthermore, the impact of multi-qubit systems on QI-scrambling has been rigorously investigated. The analysis reveals that introducing multiple qubits enhances scrambling, signifying the system's irreversible departure from its initial state over time. Intriguingly, the initial state converges the three scrambling bounds and stochastic behavior, even in the presence of multi-qubit local operators. A salient observation is the periodic nature of the scrambling behavior exhibited by the spin-star system. As the qubit count rises, the peak scrambling intensity decreases when unitary Hermitian operators are employed. The established bounds hold significant experimental promise, as they can predict both the upper and lower limits of scrambling, especially in scenarios involving multi-qubit local operators.

\section*{Acknowledgement}
This work was supported in part by National Natural Science Foundation of China(NSFC) under the Grants 12475087 and 12235008, and University of Chinese Academy of Sciences.

\section*{References}
\bibliographystyle{ieeetr}
\bibliography{bm}
 
\end{document}